\documentclass[]{aastex631}

\usepackage{cancel}

\begin{document}

\title{Velocity Dispersion of the open cluster NGC 2571 by Radial Velocities and Proper Motions\footnote{Based on observations carried out at ESO Paranal Observatory under program 096.B-0004(A)}}

\author[0000-0003-2125-8740]{Maxim V. Kulesh}
\affiliation{Ural Federal University,
51 Lenin Street, Ekaterinburg, 620000, Russia}


\author{Aleksandra E. Samirkhanova}
\affiliation{Ural Federal University,
51 Lenin Street, Ekaterinburg, 620000, Russia}

\author[0000-0002-0155-9434]{Giovanni Carraro}
\affiliation{Dipartimento di Fisica e Astronomia,
Universita' di Padova,
Vicolo Osservatorio 3, I-35122, Padova, Italy}

\author{Joao V. Sales-Silva}
\affiliation{Observat\'orio Nacional/MCTIC, R. Gen. Jos\'e Cristino, 77,  20921-400, Rio de Janeiro, Brazil}

\author[0000-0002-0155-9434]{Roberto Capuzzo Dolcetta}
\affiliation{Dipartimento di Fisica, Sapienza, Universit´a di Roma, P.le A. Moro 5, I-00165 Roma, Italy}

\author[0000-0001-8669-803X]{Anton F. Seleznev}
\affiliation{Ural Federal University,
51 Lenin Street, Ekaterinburg, 620000, Russia}



\begin{abstract}

We use a Kernel Density Estimator method to evaluate the stellar velocity dispersion in the open cluster NGC 2571.
We derive the 3-D velocity dispersion using both proper motions as extracted from Gaia DR3 and single epoch radial velocities as obtained with the instrument FLAMES at ESO VLT.
The mean-square velocity along the line-of-sight is found to be larger than the one in the tangential direction by a factor in the interval [6,8].
We argue that the most likely explanation for such an occurrence is the presence of a significant quantity of unresolved binary and multiple stars in the radial velocity sample.
Special attention should be paid to single line spectroscopic binaries (SB1) since in this case we observe the spectral lines of the primary component only, and therefore the derived radial velocity is not the velocity of the binary system center of mass.
To investigate this scenario, we performed  numerical experiments 
at varying the fractional abundance of SB1 in the observed sample.
These experiments show that the increase of the mean-square radial velocity depends actually on the fractional abundance of SB1 to a power in the range of [0.39,0.45].
We used the 3-D velocity dispersion obtained by the dispersions in the tangential directions and the assumption that the radial velocity dispersion is the same as a tangential one to estimate the virial cluster mass and the cluster mass taking into account the gravitational field of the Galaxy and the non-stationarity of the cluster.
These estimates are $650\pm30 \; M_\odot$ and  $310\pm80 \; M_\odot$, respectively, and they are in substantial agreement with the photometric cluster mass.

\end{abstract}



\section{Introduction} \label{sec:intro}

The precise determination of the mass of star clusters is an important asset for many astrophysical applications.
For instance, the cluster stellar mass is necessary to study the cluster dynamics, to evaluate whether the cluster is bound or not.
Moreover, in the case of extremely young clusters, their actual mass is key to evaluate the star formation efficiency.

It is customary to estimate the star cluster mass by two standard methods.

First, the cluster mass can be determined either through the cluster member list or by utilizing the cluster luminosity function and adopting a mass--luminosity relation \citep{NGC_4337}.
The term `photometric mass' corresponds to the mass obtained  this way.
To this aim, one needs to resort to theoretical isochrones.
One serious problem in this procedure is the difficulty to evaluate reliably the uncertainty of the resulting mass.
To overcome this obstacle, one can use the isochrone tables of different authors or to derive the error from the analytical expression for the mass-luminosity relation as given for example, by \citet{Eker+2015}.

We note that the photometric mass represents only a lower limit of the real cluster mass, because the lower end of the luminosity function, the number of invisible remnants of  massive stars and the fraction of unresolved binary and multiple stars are usually unknown \citep{Seleznev2016,Borodina+2019,Borodina+2021}.

The second method to evaluate the cluster mass relies on the knowledge of the velocity dispersion of stars in the cluster (more precisely, the mean squared of their velocity with respect to the cluster center of mass velocity).
One can use a simple formula for the `virial' mass or the formula of \citet{Danilov&Loktin2015} that takes into account the non-stationarity of the cluster and the influence of the Galactic gravitational field.
The terms `virial mass' or `dynamical mass' are typically used in this case.
This latter approach takes, in principle, into account all cluster stars, and therefore it would be an advantage with respect to the photometric estimate.
However, also this method carries some difficulties.
Firstly, one should determine some structural characteristics of the cluster, like the cluster radius, which appear in the virial equation.
To do this accurately, we need to obtain the cluster spatial density and use some numerical modeling \citep{NGC_4337}.
Secondly, for a good evaluation of the intrinsic velocity dispersion, one has to take into account the errors of the radial velocities and of the proper motions of stars.
The third problem is the contamination of the sample of radial velocities by  single-line spectroscopic binaries (SB1) since, in many cases, only one epoch radial velocity observations is available.
In this case we see only the spectral lines of the bright (primary) component of the binary and, consequently, we cannot recognize this star as a binary with single observation only.
Then, the shift of the spectral lines in the spectrum will originate both from the binary system bary-center motion and from the motion of the primary component around the bary-center.
Of course, these binaries cannot be distinguished from single stars with just single epoch observations.
Consequently, in the case of a single epoch observation, the observed radial velocity will deviate from the radial velocity of the binary system barycenter.
As a result, we would derive a radial velocity dispersion larger than the real one.
\citet{NGC_4337} had proposed that this could be a plausible interpretation of the discrepancy between the estimates of photometric and dynamical masses in the open cluster NGC 4337.

The inflation of the velocity dispersion induced by unresolved binaries in star clusters has been confirmed also by numerical N-body experiment, see for instance \citet{Rastelloetal}.\\

\noindent
In this study we propose a way to amend this bias relying on the Kernel Density Estimator (KDE; \citet{Silverman,SeleznevKDE}) statistical method and apply it to the open cluster NGC 2571.
We take the data for the proper motions of stars in this cluster from the Gaia DR3 catalog \citep{GaiaDR3} and use the original data on the radial velocities of its stars as obtained with the Very Large Telescope (VLT).
We assume for NGC 2571 distance to the Sun  1293$\pm$46 pc following the catalog of \citet{Dias+2021}.

The layout of this study is as follows:
in Section 2 we describe the observational data;
Section 3 contains the description of our method to evaluate the intrinsic velocity dispersion of stars in a cluster and its application to NGC 2571;
in Section 4 we perform numerical modelling of the sample of radial velocities with different contents of SB1 and investigate the dependence of the velocity dispersion on the fraction of SB1 systems in the sample.
Section 5 is dedicated to the mass estimates of NGC 2571.
Finally, Section 6 is devoted to the discussion of the results.

\section{Observational data} \label{sec:data}

\subsection{Astrometry}
In this study, we extract proper motion components from the Gaia DR3 catalog \citep{Gaia,GaiaDR3}.
In order to determine the tangential velocity dispersions, we use the samples of the probable cluster members derived by \citet{Cantat-Gaudin+2020} and \citet{Hunt&Reffert2023}.
Since the catalog of \citet{Cantat-Gaudin+2020} contains proper motions from the Gaia DR2 \citep{GaiaDR2}, we replaced Gaia DR3 proper motions by cross-correlating the \citet{Cantat-Gaudin+2020} list and the Gaia DR3 catalog \citep{GaiaDR3}.
For both sources, we considered as {\it bona fide} cluster members those stars having membership probability larger than 0.5.

\subsection{Radial velocities}
Radial velocities for NGC 2571 stars have been obtained using the FLAMES spectrograph onboard UT2 at ESO Paranal Observatory on the night of January 22, 2016. Observations were carried out under seeing 1.3 arcsec and thin atmospheric conditions. One 45 minutes long exposure was taken and the data were then pre-reduced using the observatory pipeline. 

We used the GIRAFFE advanced data products (ADP) processed by ESO Phase 3 that prepares and validates ESO science data. The spectra were reduced via the standard ESO GIRAFFE pipeline ({\texttt{giscience }} 2.14.2, \citet{Blecha+2000}), which consists of the bias and dark correction, localization of the fibers on the detector, flat-fielding, spectrum extraction, and wavelength calibration. After \texttt{giscience } reduction, the ESO Phase 3 performed heliocentric correction in the wavelength. We also manually inspected and cleaned the cosmic rays in each spectrum using IRAF's \texttt{splot} task. Finally, we normalized the stellar spectra through the IRAF’s \texttt{continuum} task.

We computed radial velocities and associated uncertainties of the cluster stars by cross-correlating the normalized spectra with a synthetic spectrum of similar stellar parameters from \citet{Munari+2005}.
For this, we used IRAF’s FXCOR task that implements the Fourier cross-correlation method developed by \citet{Tonry&Davis1979}.
The FXCOR task estimated the radial velocities by fitting the largest peak of the cross-correlation function, with the radial velocity determined by the peak center position. The radial velocity uncertainties were estimated using the fitted peak height and the antisymmetric noise, as described in Tonry \& Davis (1979). Finally, we obtained a list of 78 stars with the radial velocities in the field of NGC 2571.
Then, we selected stars that can be members of the cluster.
To do this, we used the mean value of the cluster proper motion from the \citet{Dias+2021} catalog and selected stars with $\mu_\alpha \in[-5.9,-3.9]$ mas/yr and $\mu_\delta \in[3.3;5.3]$ mas/yr.
These ranges for the proper motions correspond to $\pm6.2$ km/s with the assumed cluster distance of 1293 pc.
This is much larger than the typical internal mean-square velocity in open clusters.
We ended up with 34 stars having radial velocities with the proper motions in this range.
We consider these 34 stars as candidate cluster members and use them to evaluate the cluster radial velocity dispersion.

\section{Evaluation of the intrinsic velocity dispersion for NGC 2571} \label{sec:method}

\begin{figure}
   \centering
   \includegraphics[width=15truecm]{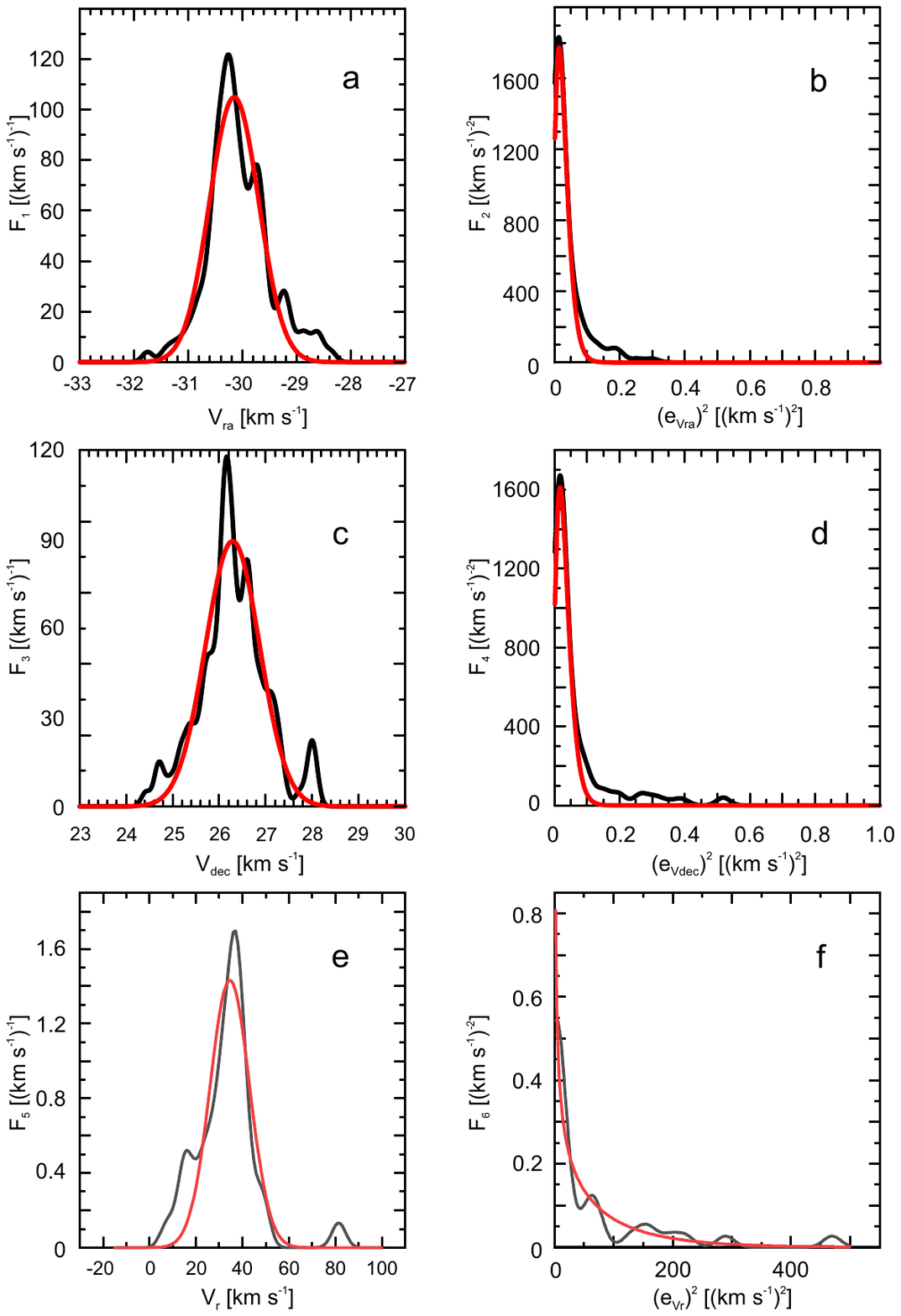}
   \caption{{\bf The left 3-panel column reports the distributions of the tangential and radial velocities (black lines) together with their fit by Gaussian functions (red lines). The right 3-panel column reports the distributions of the squared errors of velocities (black lines) and their fit by Eq. \ref{squared_errors} (red lines). Specifically: (a) Distribution of the tangential velocities in the direction of the right ascension; (b) distribution of their squared errors; (c) distribution of the tangential velocities in the direction of the declination; (d) distribution of their squared errors; (e) distribution of the radial velocities; (f) distribution of their squared errors.}}
   \label{approx}
   \end{figure}

The various formulas to derive the cluster dynamical mass (see above and below) are all based on the mean square velocity of the cluster stars.
Usually, one computes the mean square velocity in the cluster through the formula:

\begin{equation}
\label{mean-square velocity}
\langle v^2 \rangle=\frac{1}{n} \sum_{i=1}^n \left(v_i - \langle v \rangle    \right)^2 \; \mbox{,}
\end{equation}

\noindent where $\langle v \rangle$ is the velocity of the cluster center of mass, 
which we assume to be the mean velocity of the cluster stars as $\langle v \rangle$.
Note that in the above formula $v$ can be either the 3-D absolute value of the spatial velocity of the star, or its radial velocity, or a component of the tangential velocity.
The square root of $\langle v^2 \rangle$ is referred to as a velocity dispersion.

In the mathematical statistics Eq.\ref{mean-square velocity} corresponds to the biased estimate of the dispersion $\sigma^2$ which is a parameter of the Gaussian (normal) distribution.
The unbiased estimate of the dispersion $\sigma^2$ differs by a normalization factor, $(n-1)$ in the denominator instead of $n$.
In this work, we neglect this difference and consider $\langle v^2 \rangle=\sigma^2$.
Consequently, we will refer to a square root of the unbiased dispersion $\sigma^2$ also as the velocity dispersion.

The squared velocity dispersion is $\sigma^2=\sigma_r^2+\sigma_{\alpha}^2+\sigma_{\delta}^2$, where $\sigma_r$, $\sigma_{\alpha}$ and  $\sigma_{\delta}$ are, respectively, the dispersion of the radial velocities, the dispersion of the velocities in the direction of the right ascension, and the dispersion of the velocities in the direction of the declination.
Clearly, in the assumption of spherical symmetry in velocity space, the total squared velocity dispersion is $\sigma^2=3\sigma_r^2$.

Of course, with the aim of determining the velocity dispersion from Eq. \ref{mean-square velocity}, the problem of the correct accounting of the observational errors must be faced.
We propose the following method to give a reliable evaluation of the intrinsic velocity by a proper treatment of the observable distribution of velocities.

Firstly, we note that observational errors widen the distribution of velocities.
The situation is analogous to the widening of spectral line by an instrumental profile or by the joint action of the various mechanisms of spectral line broadening.
In the last case, the resulting profile results from the convolution of functions corresponding to different broadening mechanisms \citep{Gray}. 
We exploit this analogy and the properties of convolution.

If the functions describing the broadening mechanisms are all Gaussian, the result of the convolution will be a Gaussian function as well\citep{Gray}.
It is straightforward to show that in this case the dispersion of the resulting Gaussian function is the sum of the dispersions of all Gaussian functions corresponding to the broadening mechanisms.

Secondly, we plot the observable velocity distribution fitted using a Kernel Density Estimator (KDE) having a Gaussian kernel.
The KDE is a straightforward method to obtain an estimate of the distribution function, giving a continuous and differentiable estimate.
The advantages of this method as compared to the standard histogram techniques are listed in \citet{Silverman}, \citet{Merritt&Tremblay1994} and \citet{SeleznevKDE}.
Anyway, this operation, additionally broadens the observable velocity distribution.
As a result, two mechanisms concur to the broadening of the intrinsic velocity distribution: the errors and the KDE.
Then, we assume that the distribution of the velocity errors, and the intrinsic distribution of velocities are both Gaussian.
In this case, we can use the properties of  convolution (see above):

\begin{equation}
\label{dispersions}
\sigma^2_{res}=\sigma^2_{intrinsic}+\sigma^2_{errors}+\sigma^2_{KDE} \; \mbox{,}
\end{equation}

\noindent
where $\sigma^2_{res}$ is the dispersion of the resulting Gaussian function of the observed velocity distribution, $\sigma^2_{intrinsic}$ is the dispersion of the intrinsic Gaussian velocity distribution, $\sigma^2_{errors}$ is the dispersion of the Gaussian distribution of the velocity errors, and $\sigma^2_{KDE}$ is the dispersion of the Gaussian kernel. 
Clearly, the ultimate goal is to obtain $\sigma_{intrinsic}$ from the other three terms in Eq. \ref{dispersions}. Once specified the parameters of the kernel, then $\sigma^2_{KDE}$ is known.
The $\sigma^2_{res}$ is evaluated through the approximation of the resulting distribution by a Gaussian.

We cannot plot the distribution of the observational errors because we know only the absolute magnitudes of errors.
Instead, we can investigate the distribution of the squared errors.
Note that the quadratic function is a steadily increasing one.
There is a theorem in mathematical statistics which allows one to get the analytical expression for the distribution of a steadily increasing function of a random variable \citep{Taboga}.
According to it, let the random variable $\xi$ be distributed according to the density $f_{\xi}(x)$, and consider a function a steadily increasing function $g(x)$.
Then, the random variable $\eta=g(x)$ has the  density distribution $f_{\eta}(x)=(g^{-1}(x))'\cdot f_{\xi}(g^{-1}(x))$, where $g^{-1}(x)$ is an inverse function of  $g$, and $(g^{-1}(x))'$ is its derivative.
The steadily increasing function $g$ in this context is the square of the proper motion error.

We consider the observational error as a random variable having a Gaussian distribution with  mode $m$ and  dispersion $\sigma^2_{errors}$.
For the squared errors we obtain \citep{Taboga}:

\begin{equation}
\label{squared_errors}
f_{\eta}(x)=\frac{1}{2\sigma_{errors}\sqrt{2\pi x}}\cdot \exp\left(-\frac{(\sqrt{x}-m)^2}{2\sigma_{errors}^2}\right) \; \mbox{.}
\end{equation}

\noindent

We approximate the distribution of the squared errors by Eq. (\ref{squared_errors}) and then determine the $\sigma^2_{errors}$.

We applied this procedure to the distributions of tangential velocities and to the distribution of radial velocities of the probable members of the open cluster NGC 2571.
The tangential velocities are connected with the corresponding proper motions by a relation $V_t=4.74\cdot\mu\cdot r$, where $V_t$ is in km$\cdot$s$^{-1}$, $\mu$ is the proper motion in mas per year, and $r$ is the heliocentric distance in kpc.

The distributions of the velocities and their squared errors plotted by KDE are shown in Fig. \ref{approx} by black lines.
The best fits are shown by red lines: Gaussians for the velocity distributions and the functions in Eq. \ref{squared_errors} for the distributions of the velocity errors. The parameters of these approximations are summarized  in Table \ref{tab:dispres}.

One can readily see that the radial velocity dispersion is approximately 6-8 times larger than the velocity dispersion in both tangential directions.
We explain this by a pollution of the radial velocity sample by binary stars which are spectral binaries of SB1 type.
The next Section is devoted to describe the numerical experiment set up to confirm this point of view.

\begin{deluxetable*}{cccccc}
\tablenum{1}
\tablecaption{The Results of Approximation (the Standard Deviations of a Resulting Distribution and Distribution of Errors), the Standard Deviation of the Gaussian Kernel, and the Intrinsic Standard Deviation of Velocity, All in km$\cdot$s$^{-1}$\label{tab:dispres}}
\tablewidth{0pt}
\tablehead{
\colhead{Sample}&\colhead{}&\colhead{$\sigma_{res}$} & \colhead{$\sigma_{KDE}$} & \colhead{$\sigma_{errors}$} & \colhead{$\sigma_{intrinsic}$} \\
}
\startdata
Radial Velocities         & $V_r$     &   8.5$\pm$0.3   &     3          &    7.1$\pm$0.2    &   3.5$\pm$0.8   \\
Cantat-Gaudin et al.      & $V_{ra}$  & 0.454$\pm$0.005 &     0.1        & 0.0647$\pm$0.0003 & 0.438$\pm$0.005 \\
                          & $V_{dec}$ & 0.583$\pm$0.010 &     0.1        & 0.0664$\pm$0.0003 & 0.570$\pm$0.010 \\
Hunt \& Reffert           & $V_{ra}$  & 0.457$\pm$0.003 &     0.1        & 0.0868$\pm$0.0008 & 0.437$\pm$0.003 \\
                          & $V_{dec}$ & 0.558$\pm$0.005 &     0.1        & 0.0859$\pm$0.0006 & 0.542$\pm$0.005 \\
\enddata
\end{deluxetable*}

\section{The dependence of the radial velocity dispersion on the pollution of the radial velocities sample by SB1} \label{sec:modeling}

\subsection{Problem statement}
In this section we explore and evaluate the influence of SB1 stars on the radial velocity dispersion estimate.
Let us consider a binary star system with the  `primary' component mass $M$ and the `secondary' component mass $m$ defined by the mass ratio $q = m/M$.
If the spectral lines of the secondary component in the spectrum of the binary are not visible, the binary can be interpreted as  a single star.
An attempt to take into account this potential bias is now described.
 
Let us consider the relative orbit of the primary component around the secondary, see Fig.\ref{orbit_angle} for an illustration.
To make the analysis simpler, we consider the center of mass of the two stars as stationary.
The position of the primary component in its relative orbit is univocally determined by its true anomaly $\theta$.
The parameters of the orbit and the primary component position are: 

$a$ is the semi-major axis of the relative orbit;

$e$ is the relative orbit eccentricity;

$\omega$ defines the position of the periastron with respect to the line of sight;

$i$ is the inclination angle of the line of sight with respect to the orbital plane.

We then use the simple Salpeter law for the distribution of the primary component mass $M$ to model SB1 systems: 
$$F(M) \sim M^{-2.35}, \; M\in\left[0.1, 10\right] M_\odot \; .$$

We take the distribution of $q$ from our recent study \citet{Malofeeva+2022}:
$$F(q) \sim q^{-0.53}, q\in\left[0.1, 0.9\right]$$.

As for the distribution of the semi-major axes $a$, we consider it uniform in the range of $\left[10, 3\times10^5\right]$ AU \citep{Tutukov&Kovaleva2019}.

The so-called {\it thermal} distribution is assumed for the distribution of eccentricities $e$ \citep{Jeans1919}: $F(e) \sim e$.
The distributions of all angles are uniform: $i$ in the range of $[0, \pi]$, $\omega$ and $\theta$ in the range of $[0, 2\pi]$.

At this point, the fundamental question to answer to is: what is the distribution of the primary component line-of-sight velocity relative to the center of mass of the binary system?

\subsection{Solution}

\begin{figure}
    \centering
    \includegraphics[width=12truecm]{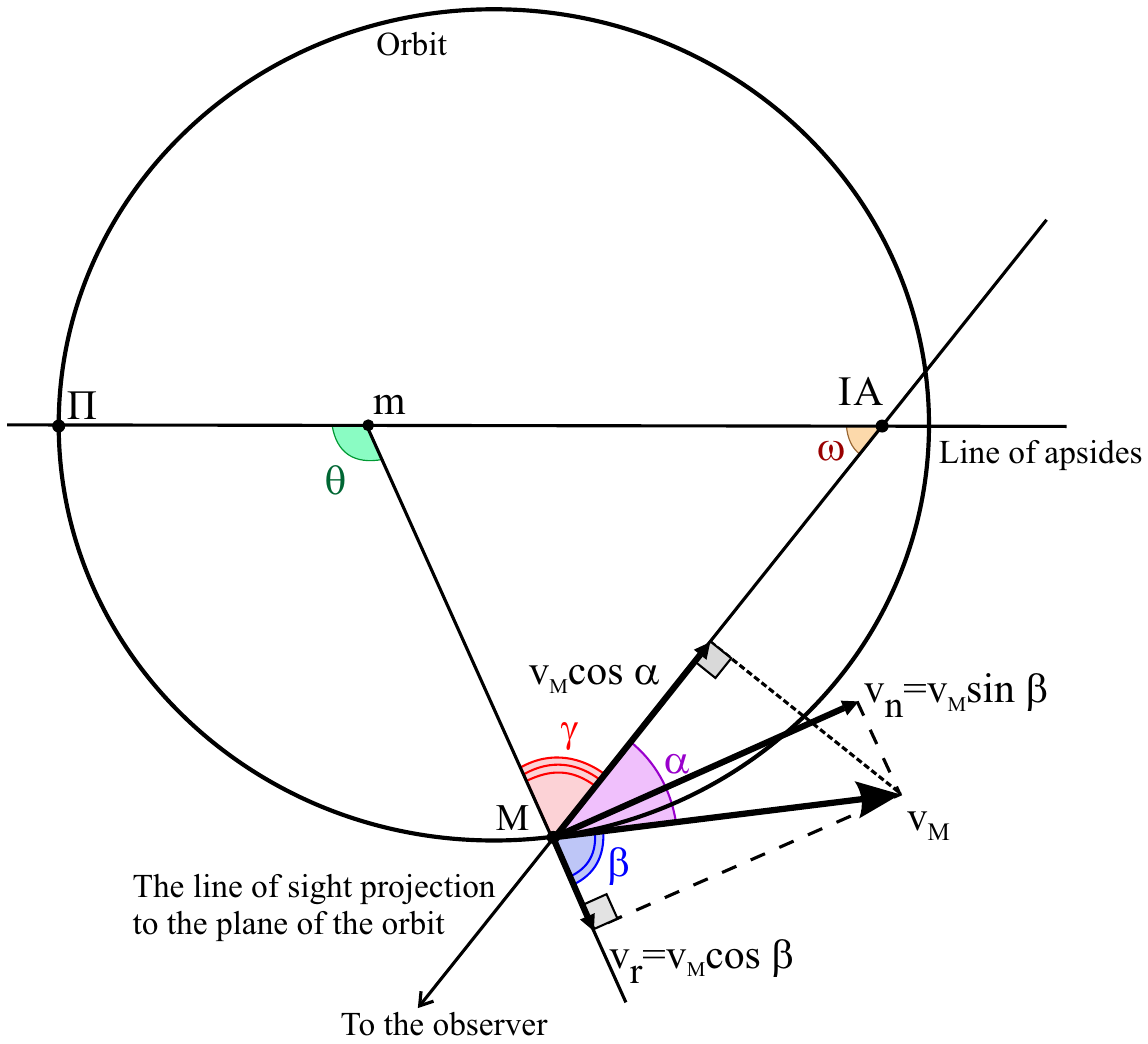}
    \caption{Orbit of the primary component $M$ relative to the secondary component $m$. $\Pi$ is the periastron. $IA$ is the intersection point of the line of sight projection and the line of apsides. $\theta$ is the true anomaly.}
    \label{orbit_angle}
\end{figure}

To provide an answer to this question, we consider the case when the line of sight is not perpendicular to the orbital plane.
Fig.\ref{orbit_angle} sketches the orbit of the primary component $M$ relative to the secondary component $m$ and the projection of the line of sight onto the orbital plane. 

The primary component M moves along the orbit with the semi-major axis $a$ and eccentricity $e$ in accordance with the solution of the unperturbed two-body problem with the full velocity $\mathbf{v_M}$.
We should find the component of this velocity along the line of sight.
Let $\alpha$ be the angle between the vector $\mathbf{v_M}$ and the projection of the line of sight onto the orbital plane.
Let $\Pi$ be the orbit periastron, hence $\theta$ is the true anomaly of $M$. 

Let $IA$ be the intersection point of the line of sight projection and the line of apsides, then $\omega$ is the angle between the line of sight projection and the line of apsides.
It is clear from Fig.\ref{orbit_angle} that the angle between the radius-vector of the primary component and the projection of the line of sight onto the orbital plane is
\begin{equation}
    \gamma=\theta-\omega
    \label{gamma}
\end{equation}

The orientation of the primary component velocity vector $\mathbf{v_M}$ relative to the line of sight projection can be calculated with its radial ($v_r$) and tangential ($v_n$) components relative to the primary component radius vector.
Let $\beta$ is the angle between the primary component radius vector and the primary component velocity.

The $v_r$ and $v_n$ values depend on the true anomaly $\theta$ and eccentricity $e$:

\begin{equation}
    v_r(\theta)=\frac{v_{c}}{\sqrt{1-e^2}}e\sin \theta,
    \label{vR}
\end{equation}

\begin{equation}
    v_n(\theta)=\frac{v_{c}}{\sqrt{1-e^2}}\left(1+e\cos \theta\right),
    \label{vn}
\end{equation}

\noindent where the circular velocity $v_c$ depends on $M$, $q$, $a$ and on the gravitational constant $G=887.125$~AU$\times$(km/s)$^2\times M_\odot^{-1}$ in appropriate units:

\begin{equation}
    v_c=\sqrt{\frac{GM(1+q)}{a}}.
    \label{vc}
\end{equation}

We obtain $\tan \beta$ dividing by mean of Eq.\ref{vn} and Eq.\ref{vR}, so we can `````````````````````````````````
evaluate $\beta$ as:

\begin{equation}
    \beta=\arctan \frac{1+e\cos\theta}{e \sin \theta} =\arctan \frac{\frac{1}{e}+\cos\theta}{\sin \theta}.
    \label{beta}
\end{equation}

From Fig. \ref{orbit_angle} one can readily see that $\alpha + \beta + \gamma = \pi$.
Thus, by substituting Eq.(\ref{gamma}) we obtain:

\begin{equation}
    \alpha = \pi - \beta - \gamma = \pi - \beta - \theta + \omega
    \label{alpha}
\end{equation}

The value of $v_M$ for any parameters can be inferred from the {\it vis-viva} equation for any $v_c$, $a$ and the radius-vector modulus $r$:

\begin{equation}
    v_M^2(r)=v^2_c \left(\frac{2a}{r}-1\right),
    \label{visviva}
\end{equation}

$r$ is linked to  $\theta$, $a$, and $e$ trough the orbit equation (equation of an ellipse) in polar coordinates:

\begin{equation}
    r(\theta) = \frac{a\left(1-e^2\right)}{1+e\cos\theta} \; .
    \label{polar}
\end{equation}

We can evaluate $v_M(\theta)$ by plugging Eq.(\ref{polar}) into Eq.(\ref{visviva}):

\begin{equation}
    v_M(\theta) = v_c \sqrt{\frac{2(1+e\cos\theta)}{1-e^2}-1}=v_c \sqrt{\frac{e^2+2e\cos\theta+1}{1-e^2}}
    \label{vM}
\end{equation}

since we know both $v_M(\theta)$ and $\alpha$.
This way we can find the projection of the primary component velocity onto the projection of the line of sight on the orbital plane:

\begin{equation}
    v_1 = v_M(\theta) \cos\alpha \; .
    \label{cosalpha}
\end{equation}

\begin{figure}
    \centering
    \includegraphics[width=8truecm]{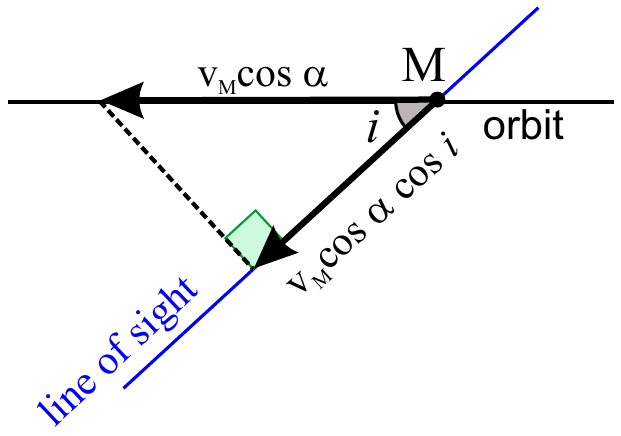}
    \caption{The plane passing through the primary component, containing the line of sight and perpendicular to the orbital plane. The angle $i$ is an inclination of the line of sight to the orbital plane. $v_M \cos \alpha \cos i$ is the desired projection of the velocity of the main component onto the line of sight.}
    \label{line_i}
\end{figure}

\noindent
Finally, the desired projection of the velocity of the main component onto the line of sight is (see Fig.\ref{line_i}):

\begin{equation}
    v_2 = v_1 \cos i = v_M \cos \alpha \cos i
    \label{inclin}
\end{equation}

\begin{figure}
     \centering
     \includegraphics[width=12truecm]{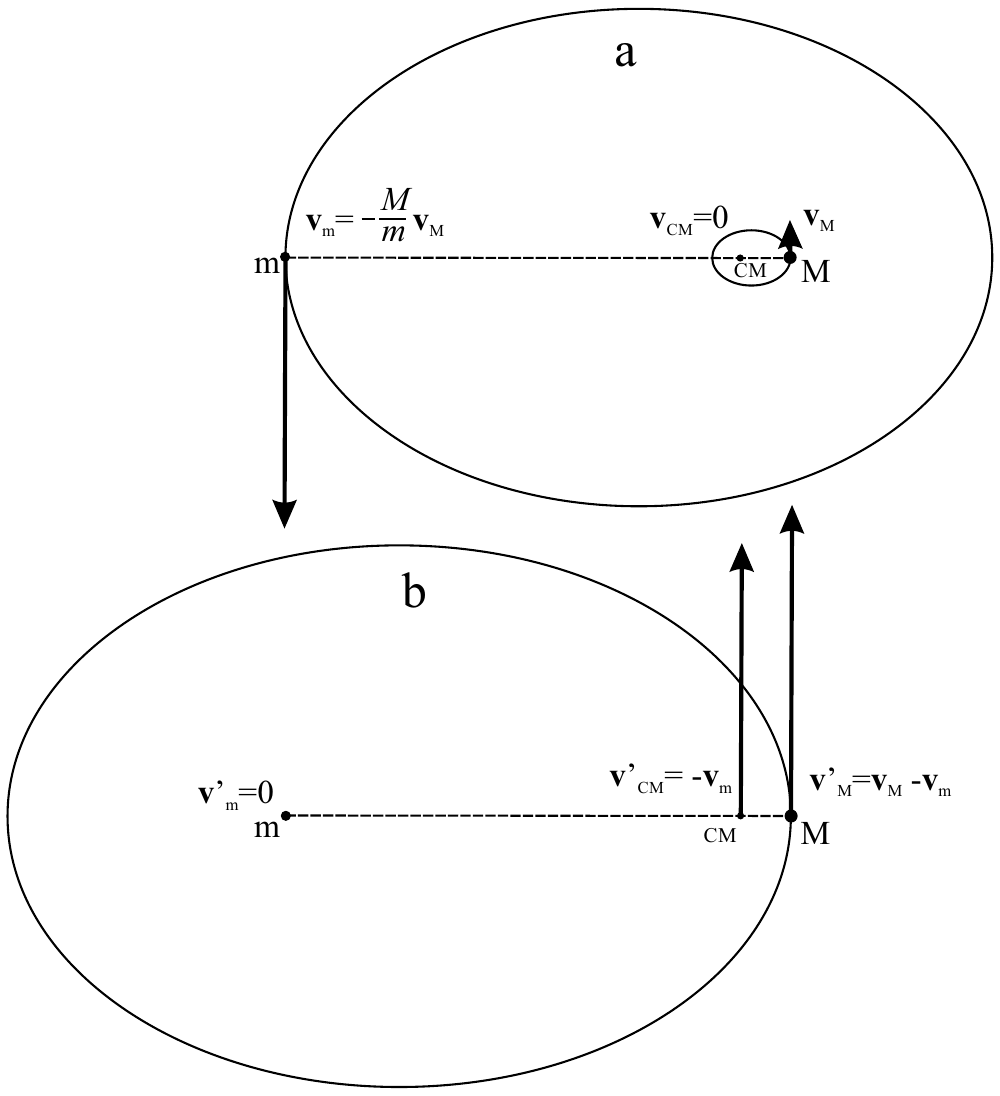}
     \caption{(a) Velocities of main component $M$ and secondary component $m$ in a coordinate system associated with the center of mass $CM$, $\mathbf{v_m}=-(M/m)\mathbf{v_M}$ by conservation of momentum.
     (b) Velocities of main component $M$ and center of mass $CM$ in a coordinate system associated with secondary component.
     }
     \label{orbit_CM}
     \end{figure}
\noindent
Now we must move to a reference frame associated with the center of mass of the binary system.
For convenience, we perform this reference frame change in reverse order.
Let us consider the coordinate system where center of mass $CM$ is stationary, then the primary component $M$ has velocity $\mathbf{v_M}$, while the secondary component $m$ has velocity $\mathbf{v_m}$.
The total momentum of the system is equal to zero, so we can evaluate one velocity from another:

\begin{equation}
    M\mathbf{v_M}+m\mathbf{v_m}=\mathbf{0},\ \mathbf{v_m}=-\frac{1}{q}\mathbf{v_M}
\end{equation}

\noindent
To move to a reference frame with a stationary secondary component, we should subtract $\mathbf{v_m}$ from the velocity of all points of the system.
Then, the new velocity of the primary component $\mathbf{v'_M}$ is:

\begin{equation}
    \mathbf{v'_M} = \mathbf{v_M}-\mathbf{v_m}= \left(\frac{q+1}{q}\right)\mathbf{v_M}
\end{equation}

Therefore in order to get the velocity in the reference frame with stationary $CM$ from the velocity on the orbit relative to the secondary component, we should multiply this velocity by a factor of $q/(1+q)$.
As a result, we obtain the velocity of the primary component relative to the center of mass $CM$.
In other words, this velocity will be an addition to the velocity of the binary system center of mass in case of single observation of SB1:

\begin{equation}
    v_{SB1} = v_2 \frac{q}{1+q}
    \label{qq}
\end{equation}

Summarizing, we evaluate the radial velocity of one SB1 system following this procedure:

\begin{enumerate}
\item Initialize parameters $q$, $M$, $a$, $e$, $\omega$, $i$ and $\theta$ from their corresponding distributions (see above).
\item Calculate the supplementary angle $\beta=\arctan \left[ \left( 1/e+\cos\theta \right) / \sin \theta \right]$ (see Eq.(\ref{beta})).
\item Calculate supplementary angle $\alpha=\pi - \beta - \theta + \omega$ (see Eq.(\ref{alpha})).
\item Calculate circular velocity $v_c=\sqrt{GM(1+q)/a}$ (see Eq.(\ref{vc})).
\item Calculate total velocity $v_M=v_c \sqrt{(e^2+2e\cos\theta+1)/(1-e^2)}$ (see Eq.(\ref{vM})).
\item Calculate line-of-sight velocity $v_{SB1} = v_M q \cos \alpha \cos i /(1+q)$ (see Eqs.(\ref{cosalpha}), (\ref{inclin}) and (\ref{qq})).
\end{enumerate}

\subsection{The influence of SB1 systems on the cluster velocity dispersion}

\begin{figure}
     \centering
     \includegraphics[width=15truecm]{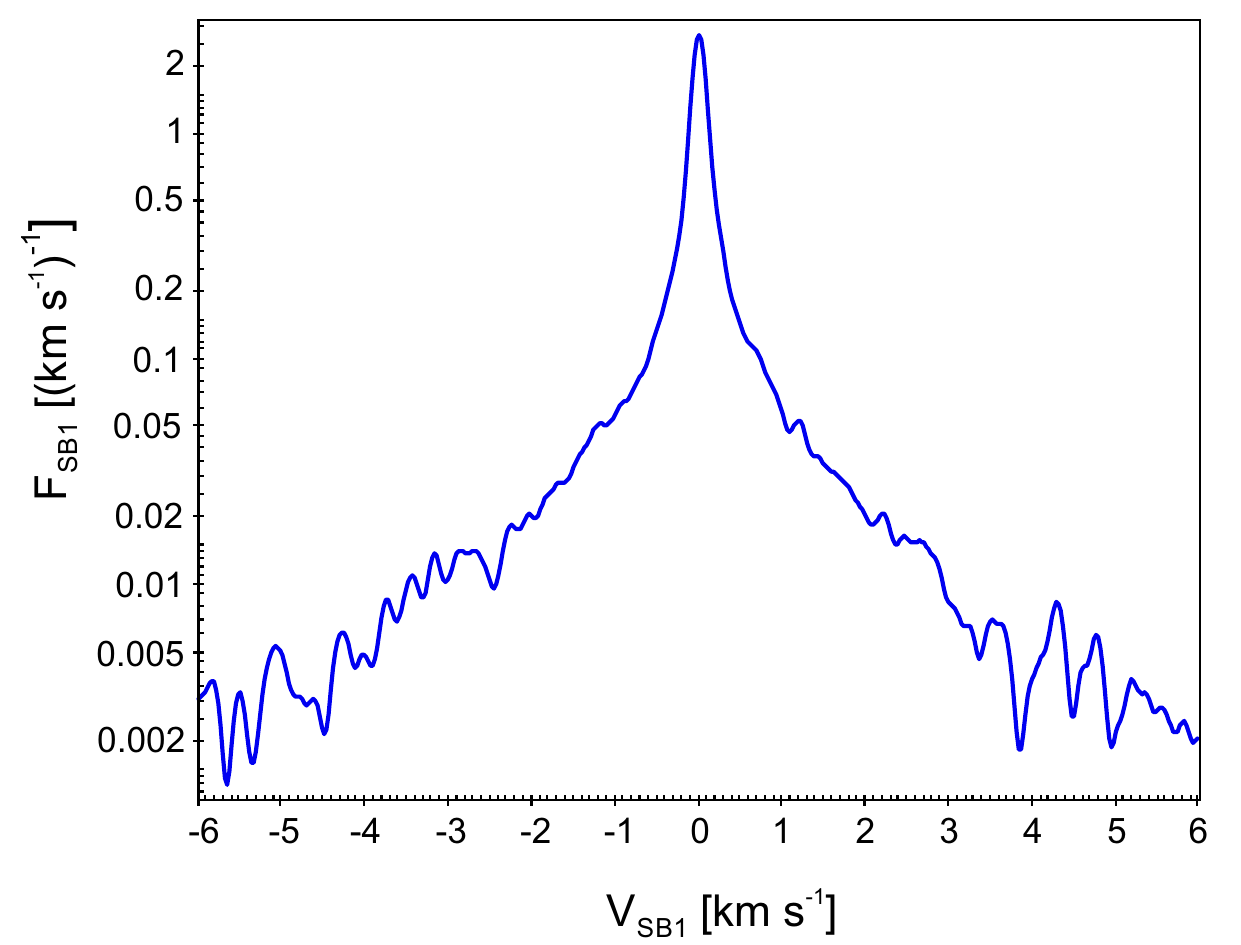}
     \caption{Distribution of the primary component radial velocities in SB1 systems relative to the center of mass of the binary system in the logarithmic scale.}
     \label{distr_SB1}
\end{figure}

Firstly, we estimate the distribution of the primary component radial velocities in SB1 systems relative to the center of mass of the binary system.
We achieve this by generating randomly a large number (100~000) of input parameters of such systems that satisfy the input distribution laws.
The result is shown in Fig.\ref{distr_SB1}.
The distribution is very narrow with wide tails.
 $V_{SB1}$ is the difference between the line-of-sight projection of the primary component of the binary system and the radial velocity of the binary system mass center.
In other words, it turns into an error which is introduced, if we use the radial velocity of SB1 considering it as a single star (note that we can not distinguish SB1 with a  single epoch observation).
This error results in the overestimation of the radial velocity dispersion of the cluster. 

Secondly, we answer the question: how does an additional SB1 radial velocity modify the cluster's radial velocity standard deviation ($\sigma$) ?
To do this, we consider a simple cluster model of $N=5000$~stars with standard deviation of radial velocity  $\sigma_0=3^{-1/2}\approx 0.58$~km/s.
Every star is represented by its radial velocity $v_r$, generated from normal distribution with the mean $\mu=0$~km/s and $\sigma=\sigma_0$.
Then we specify the fraction of SB1 stars $f$ among the cluster members.
For every SB1 star, we add the additional velocity $v_{SB1}$, generated from the distribution $F_{SB1}$ as shown in Fig.\ref{distr_SB1}.
Taking these new velocities into account, we determine new value of the mean-square velocity of the cluster model.

\begin{figure}
     \centering
     \includegraphics[width=15truecm]{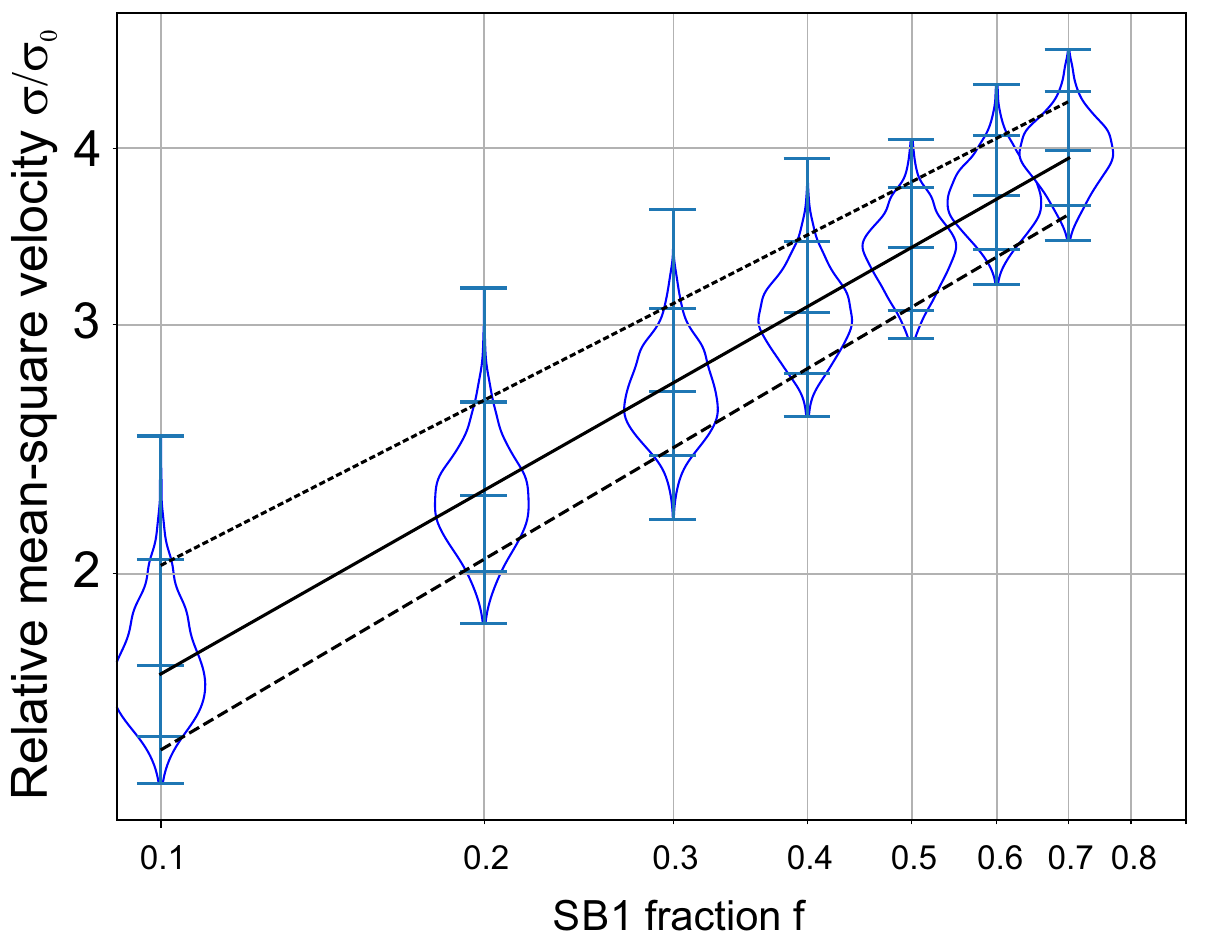}
     \caption{Violin plot for the dependence of the relative mean-squared velocity $\sigma/\sigma_0$ on the SB1 fraction $f$ in a square-root scale.
     The solid line is an approximation of the 50\% quantile, the small-dashed line is an approximation of the 5\% quantile, the dashed line is an approximation of the 95\% quantile.}
     \label{sigma_from_frac}
\end{figure}

By generating around $1000$ of such cluster models for every value of $f$, we get a violin plot for the dependence of the relative mean-square velocity $\sigma/\sigma_0$ on the SB1 fraction $f$.
Figure \ref{sigma_from_frac} shows this plot in a square-root scale for seven values of the relative mean-square velocity.
The horizontal lines at the vertical bars shows 0\%, 5\%, 50\%, 95\%, and 100\% quantiles of the distribution.
We fitted quantiles 5\%, 50\%, and 95\% with the following function:

$$\log \frac{\sigma}{\sigma_0} = a \log f + b \; ,$$

\noindent where $\sigma_0=0.58$~km/s (see above), and $\sigma$ is the root mean-square velocity obtained after the addition of the fractional abundance $f$ of SB1 binaries.
$\sigma_l$ and $\sigma_u$ correspond to quantiles 5\%, and 95\%, respectively.
Th results are as follows:

$$\log \frac{\sigma}{\sigma_0} = 0.43 \log f + 0.66 \; ,$$
$$\log \frac{\sigma_l}{\sigma_0} = 0.45 \log f + 0.62 \; ,$$
$$\log \frac{\sigma_u}{\sigma_0} = 0.39 \log f + 0.70 \; .$$

As expected, the relative mean-square velocity $\sigma/\sigma_0$ grows with $f$
If the velocity distribution of the primary components of SB1 binaries were  Gaussian one, one would expect:

$$\sigma^2 = \sigma_0^2+f\sigma_{SB1}^2 \; ,$$

\noindent 
in agreement with the properties of the convolution of two Gaussian distributions.
In that case the dependence of the relative mean-square velocity on the fractional abundance $f$ of SB1 binaries would be:

$$\frac{\sigma}{\sigma_0} = \sqrt{1+f\left(\frac{\sigma_{SB1}}{\sigma_0}\right)^2} \sim \sqrt{f} \; .$$

\noindent 
However, in our case, the exponent in the dependence of the relative mean-square velocity on the fractional abundance $f$ of SB1 binaries is less than 0.5 due to non-Gaussian nature of the velocity distribution of the primary components of SB1 binaries (see Fig.\ref{distr_SB1}).

\section{The mass of NGC 2571} \label{sec:mass}

We used the velocity dispersion in the tangential direction to estimate the dynamical mass of NGC 2571.
We assume that the velocity distribution in the line-of-sight direction is the same as the velocity distribution in both tangential directions and evaluate the squared velocity dispersion as $\sigma^2=(3/2)(\sigma_{\alpha}^2+\sigma_{\delta}^2)$.

To estimate the cluster mass, we follow to \citet{NGC_4337}, Section 9.2.
We estimate the virial cluster mass by the formula:

\begin{equation}
\label{virial mass}
M_{vir}=\frac{2\sigma^2\bar R}{G} \; \mbox{,}
\end{equation}

\noindent 
and the cluster mass taking into account the gravitational field of the Galaxy and the cluster non-stationarity by the formula of \citet{Danilov&Loktin2015}:

\begin{equation}
\label{mass_dan}
M_d=\frac{2\bar RR_u\left[2\sigma^2-\displaystyle{\frac{(\alpha_1+\alpha_3)\langle r^2 \rangle}{3}}\right]}{G(\bar R+R_u)}  \; \mbox{,}
\end{equation}

\noindent where $R_u=\langle 1/r_i \rangle ^{-1}$ is the mean inverse star distance to the cluster centre, $\bar R=\langle 1/r_{ij}\rangle ^{-1}$ is the mean inverse star-to-star distance, $\langle r^2\rangle $ is the mean square of the star distance to the cluster centre, $\alpha_1$ and $\alpha_3$ are the field constants \citep{Chandrasekhar1942} characterising the Galactic potential, $\Phi(R,z)$ in Galacto-centric cylindrical coordinates, in the vicinity of a star cluster:

\begin{equation}
\label{alpha1}
\alpha_1=\left(\frac{1}{R}\frac{\partial\Phi}{\partial R}-\frac{\partial^2\Phi}{\partial^2 R}\right)_{R_{cl}}=4A(B-A)<0  \; \mbox{,}
\end{equation}

where $A$ and $B$ are the Oort's constants, and

\begin{equation}
\label{alpha3}
\alpha_3=-\left(\frac{\partial^2\Phi}{\partial^2 z}\right)_{z_{cl}}>0  \; \mbox{.}
\end{equation}

\noindent
$R_{cl}$ and $z_{cl}$ are the cluster center of mass cylindrical coordinates.
The values of $\alpha_1$ and $\alpha_3$ were calculated adopting the Galactic potential model of \citet{Kutuzov&Osipkov}. Arguments in favour of this model are listed in \citet{Seleznev2016}.

We performed the following steps to evaluate the cluster mass by Eq.\ref{virial mass} and Eq.\ref{mass_dan}.

\begin{figure}
     \centering
     \includegraphics[width=12truecm]{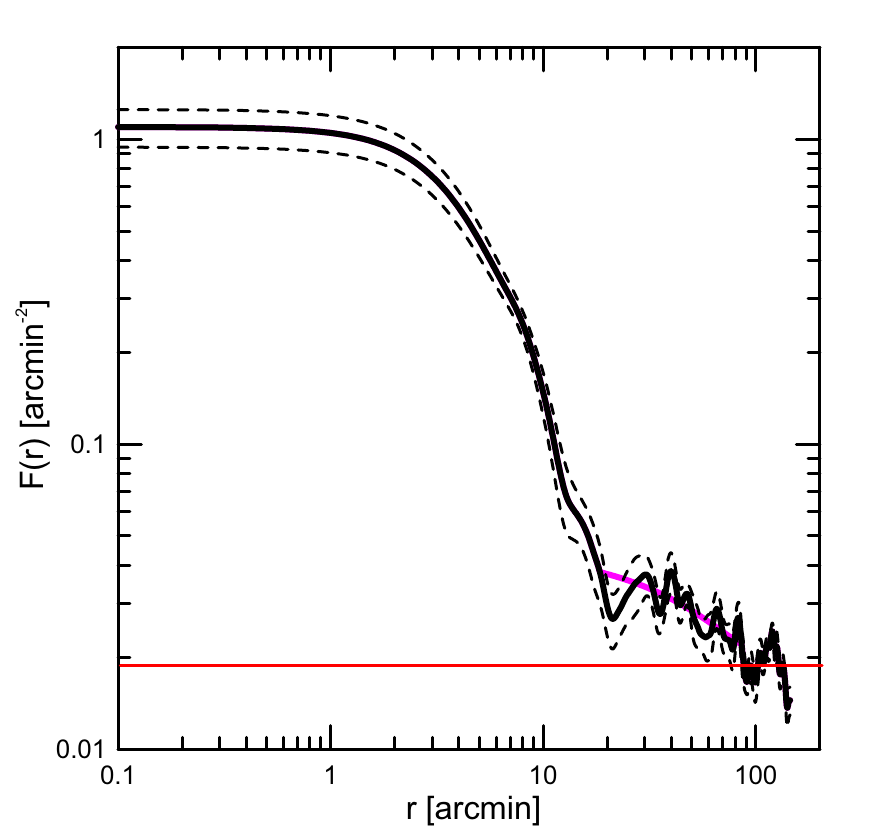}
     \caption{The surface density profile of NGC 2571 --- solid black line.
     Dashed black lines show the confidence interval.
     Horizontal red line shows the mean field density.
     The smooth magenta line at bottom right is a polynomial approximation of the surface density in the cluster corona.}
     \label{profile}
\end{figure}

1. First we  determined the cluster radius following the method proposed in \citet{Seleznev2016}.
This method is based on the comparison of the surface density radial profile of the cluster with the mean density of surrounding field stars.
The cluster radius is the distance from the cluster center where the radial density profile  intersects the (horizontal) line of the mean density of field stars (see \citet{Seleznev2016} for more details).
To this aim we have selected stars from Gaia DR3 catalog \citep{GaiaDR3} in a  field 5$\times$5 degrees around the cluster center in the parallax and proper motion range of $\varpi\in[0.5;1.0]$ mas, $pmra\in[-5.4;-4.5]$ mas/yr and $pmdec\in[3.8;4.7]$~mas/yr with $G\leqslant18$ mag, respectively, to minimize the field stars contamination.
The cluster radius results to be $R_c=87.9\pm1.3$ arcmin.
Then, the cluster is found to have a large corona with low stellar density.
All this is  illustrated in Fig.\ref{profile} where the cluster radial density profile (solid black line) shows the cluster corona extending from approximately 20 arcmin to nearly 90 arcmin.
The outer radius of the cluster corona (the cluster radius) is determined as the first intersection of the radial density profile with the line of the mean density of the field stars (horizontal red line).
This fact is confirmed by the map of the whole of \citet{Hunt&Reffert2023} sample.
This sample has an extended asymmetric corona of stars with the membership probability less than 0.5.

2. We have fitted the surface density $F(r)$ in the cluster corona region by a second order polynomial and obtained steadily decreasing function.

3. We obtain the spatial distribution of stars $f(r)$ around the cluster center using the solution of the Abel equation proposed by \citet{vonZeipel&Lindgren1921}:

\begin{equation}
\label{spadens}
f(r)=\frac{1}{\pi}\int \limits_0^{\sqrt{R_c^2-r^2}}S(\sqrt{r^2+z^2})dz \; \mbox{,}
\end{equation}

\noindent where

\begin{equation}
\label{spadens2}
S(r)=-\frac{1}{r}\frac{dF(r)}{dr} \; \mbox{.}
\end{equation}

4. To estimate the values of $\bar R$, $R_u$ and $\langle r^2\rangle$, we have performed a Monte Carlo sampling of the spatial density profile $f(r)$.
Twenty different Monte Carlo samples were built, in order to estimate the scatter in the estimates.
We obtained: $\bar R=1.90\pm0.08$ pc, $R_u=0.60\pm0.21$ pc, and $\langle r^2\rangle=5.39\pm0.31$ ${\rm pc^2}$ for a cluster distance of 1293 pc \citep{Dias+2021}.

5. Finally, we have obtained $M_{vir}=650\pm30 \; M_\odot$ and $M_d=310\pm80 \; M_\odot$
using the velocity dispersion for the \citet{Cantat-Gaudin+2020} sample of member stars.
Masses corresponding to \citet{Hunt&Reffert2023} sample differ negligibly.

It is interesting to compare these estimates with the cluster photometric mass.
We took the sample of the NGC 2571 probable members from \citet{Cantat-Gaudin+2020} and the PARSEC isochrone \citep{PARSEC} of the age and metallicity corresponding to the cluster data from \citep{Dias+2021}.
The stellar absolute magnitudes were obtained adopting  distance and reddening from \citet{Dias+2021}.
The lower estimate of the cluster photometric mass turned out to be $M_{ph}=290\pm20 \; M_\odot$ for probable cluster members with the magnitude $G\leqslant18$ mag (this magnitude corresponds to approximately 0.7 $M_\odot$).
The uncertainty corresponds to the cluster distance and reddening errors from \citet{Dias+2021}.
Also, we took into account the uncertainty of the mass obtained from the theoretical isochrone {\bf \citep{PARSEC}} to be 0.11 $M_\odot$.

The cluster photometric mass is in a rough agreement with the estimates of the cluster dynamical mass considering that we are neglecting  weak invisible stars, dark remnants, and unresolved binary and multiple stars (see the analysis in \citet{Seleznev2016,NGC_4337} and \citet{Borodina+2021}).
We can evaluate the total cluster mass considering all these populations.
The \citet{Cantat-Gaudin+2020} sample has magnitudes in the range $G\in[8.7;18]$.
This magnitude range corresponds to the mass range of $m\in[0.71;7.18] \; M_\odot$ considering the cluster distance and reddening from \citet{Dias+2021} and the theoretical isochrone of \citet{PARSEC} for the cluster age $\log t=7.52$ \citep{Dias+2021}.
We assume that the total range of the stellar mass in the cluster at the beginning was $m\in[0.08;60] \; M_\odot$ and use the \citet{Kroupa2001} initial mass function.

We consider that stars with an initial mass less than 8 $M_\odot$ evolve to white dwarfs \citep{Limongi+2023} with a mean mass of $0.79\pm0.16 \; M_\odot$ \citep{Suleimanov+2019}.
Stars with the initial mass $m\in[8;27] \; M_\odot$ and solar metallicity explode as Supernovae \citep{Heger+2003} and leave neutron stars with the mean mass of $1.4 \; M_\odot$.
Stars with  initial mass $m\in[27;65] \; M_\odot$ and solar metallicity also explode as Supernovae \citep{Heger+2003}, but leave  black holes with a relatively small mass  \citep{Heger+2003}.
The latter evolutionary path is the most ambiguous since the exact mass loss is unknown.
For the sake of simplicity, we assume that the mass of the black hole is equal to one half of the progenitor's mass.

Finally, we take into account the presence of unresolved binary and multiple stars following to scheme described in \citet{Borodina+2021}.
As for the fractional abundance of the unresolved binary and multiple stars, we assume it to be similar as in the Alpha Persei cluster $\alpha=0.48$ \citep{Malofeeva+2022}, which is the closest in age  among the four clusters studied in \citet{Malofeeva+2022}).
According to \citet{Borodina+2021} the cluster mass increment is approximately 1.2.
This yields a total cluster mass estimate of $700\pm200\; M_\odot$.

We have estimated the spread of the total cluster mass estimate by a numerical experiment.
To do this we performed the estimate of the total cluster mass (see above) 1000 times varying the exponents of the \citet{Kroupa2001} mass function and the {\it visible} cluster mass ($M_{ph}=290\pm20 \; M_\odot$).
All varying values were considered as distributed by a Gaussian law, their errors were used as a standard deviation.

The resulting distribution is skewed (with a long tail towards large values of mass) with a mode of approximately 700 $M_\odot$.
As associated error we assume the standard deviation obtained after the removal of values exceeding the mode by three times the initial standard deviation.

The estimate of $700\pm200\; M_\odot$ has been derived adopting the \citet{Kroupa2001} initial mass function, but the cluster actual total mass is expected to be somewhat smaller due to dissipation during the relaxation phase.

In conclusion, this photometric estimate of the total cluster mass is in a good agreement with the estimate of the cluster dynamical mass.

\section{Summary and conclusions} \label{sec:summary}

In this study we present a method to estimate the mean square velocity of  star cluster probable members from radial velocities and proper motions measurements, taking into account the errors of these values.
To do this we assume that the errors follow a normal distribution and use the properties of the convolution of several normal distributions.

In particular, we consider the mean square velocity of the probable members of the open cluster NGC~2571 obtained from radial velocities obtained at the VLT telescope and from Gaia DR3 proper motions.
The mean square velocity considering radial velocities only (line of sight mean square velocity) turns out to be 6-8 times larger than the mean square velocity derived from proper motion components only (the tangential mean square velocity).
We have shown that the larger value of the line of sight  mean square velocity can be explained by the pollution of the sample with the radial velocities of spectroscopic binaries with spectral lines of the primary component (SB1) only.
These binaries cannot be distinguished from single stars with just a single epoch observations.
Figure \ref{sigma_from_frac} shows that a fractional abundance, $f$, of SB1 binaries of just 0.2 leads to an overestimation of the mean square velocity of a factor around 2.3.
In turn, the squared mean square velocity (and hence the cluster virial mass) will be overestimated of about 5.3 times.

In the specific case of NGC 2571, we have estimated both the virial mass and the mass resulting from the formula of \citet{Danilov&Loktin2015} using the mean-squared velocity from the proper motions only.
Both values are in substantial agreement with the photometric mass, derived accounting for the cluster probable members.

\citet{LiLu+2020} showed that proper motions also can be influenced by  unresolved binaries.
However, this influence seems to be negligible given the substantial agreement between NGC 2571 photometric and member-based dynamical mass.

Our results demonstrate that one can use radial velocities to estimate the mass of a star cluster only after carefully cleaning for spectroscopic binaries.

\begin{acknowledgments}
This work has made use of data from the European Space Agency (ESA) mission {\it Gaia} (\url{https://www.cosmos.esa.int/gaia}), processed by the {\it Gaia} Data Processing and Analysis Consortium (DPAC, \url{https://www.cosmos.esa.int/web/gaia/dpac/consortium}).
Funding for the DPAC has been provided by national institutions, in particular the institutions
participating in the {\it Gaia} Multilateral Agreement.
The work of M.V.Kulesh and A.F.Seleznev was supported by the Ministry of Science and Higher Education of the Russian Federation by an agreement FEUZ-2023-0019.
\end{acknowledgments}

\bibliography{NGC_2571}{}
\bibliographystyle{aasjournal}



\end{document}